\documentclass{PoS}
\usepackage{xspace}

\title{Cosmic particles}

\ShortTitle{Cosmic particles}

\author{\speaker{Mariangela Settimo}
        Laboratoire de Physique Nucl\'eaire et de Hautes Energies (LPNHE), CNRS/IN2P3, Paris, France\\
        E-mail: \email{mariangela.settimo@lpnhe.in2p3.fr}}

\abstract{Since more than a century we investigate cosmic particles coming from the Universe with the aim of understanding their nature, their origin and how they are accelerated. So far, cosmic rays have provided many impressive results, giving birth to the particle physics and extending our vision of the Universe over a wide range of energies.  In this contribution an overview of the most recent result is given. This field remains extremely active with new experiments and instruments opening new perspective for the research at the edge between astrophysics and particle physics.}

\FullConference{38th International Conference on High Energy Physics\\
		3-10 August 2016\\
		Chicago, USA}

\begin{document}

\section{Cosmic rays: a vast field}

Since their discovery, more than hundred years ago, cosmic particles have delivered major results, giving birth to particle and high-energy physics and later on to the development of the different branches of astroparticle physics.  The observation of these particles has opened new perspectives on the study of our Universe and on the most violent phenomena known so far and it may offer a unique chance to test fundamental physics in regimes not accessible to laboratory-based experiments. 
Cosmic rays are particles arriving from outside the atmosphere, such as photons, neutrinos, electrons and charged nuclei. The nature, the flux intensity and the arrival direction of these particles are the main observables to understand their sources, their acceleration mechanisms and their propagation in the galactic and extra-galactic environment.
This short proceeding aims to provide a review of some recent results with a main focus on charged cosmic rays. 

The most striking feature of cosmic rays is that the spectrum extends for more than 11 orders of magnitude in energy $-$ from 10$^{9}$~eV to a few 10$^{20}$~eV $-$ with a flux varying from a few particles per m$^2$ per second at the lowest energies to one particles per km$^2$ per year above 10$^{19}$~eV. 
Cosmic rays in the low energy region - having a Larmor radius small compared to the size of the galactic disk -  are of galactic origin and supernova remnants (SNR) are considered the principal candidate sources.   
At the highest energies, cosmic rays are supposed to be accelerated in extragalactic sources which are not yet identified (among them for example the Active Galactic Nuclei, AGN).  The transition between galactic and extragalactic cosmic rays is expected to occur around 10$^{17} - 10^{18}$~eV and remains an important question to solve. 

In a first approximation the spectrum of cosmic rays is well described by a single power-law, $E^{-\Gamma}$, with $\Gamma \sim$~2.7 with the exception of a few breaks.  
Three predominant spectral changes have been identified: (i) A steepening, the ``knee", at energies of a few 10$^{15}$~eV, which is commonly interpreted as due to the maximum acceleration of protons in galactic sources~\cite{knee}; (ii) A hardening at energies around 10$^{18}$~eV, named "ankle", historically linked to the transition from a steeper galactic component to a flatter extragalactic one or interpreted as a proton propagation effect~\cite{DipModel}; A flux suppression at energies above 10$^{19.5}$~eV whose nature, still under debate,  may be connected to the predicted GZK cut-off or to a limitation in the maximum power of the sources\cite{GZK1,GZK2,Allard,Allard2}. 
The large statistics and small systematic uncertainties achieved by the new experiments is allowing the observation of sub-leading structures which are important keys to constrain the current models. 

\begin{figure}[!t]
\hspace{-0.5cm}
\includegraphics[width=0.56\textwidth]{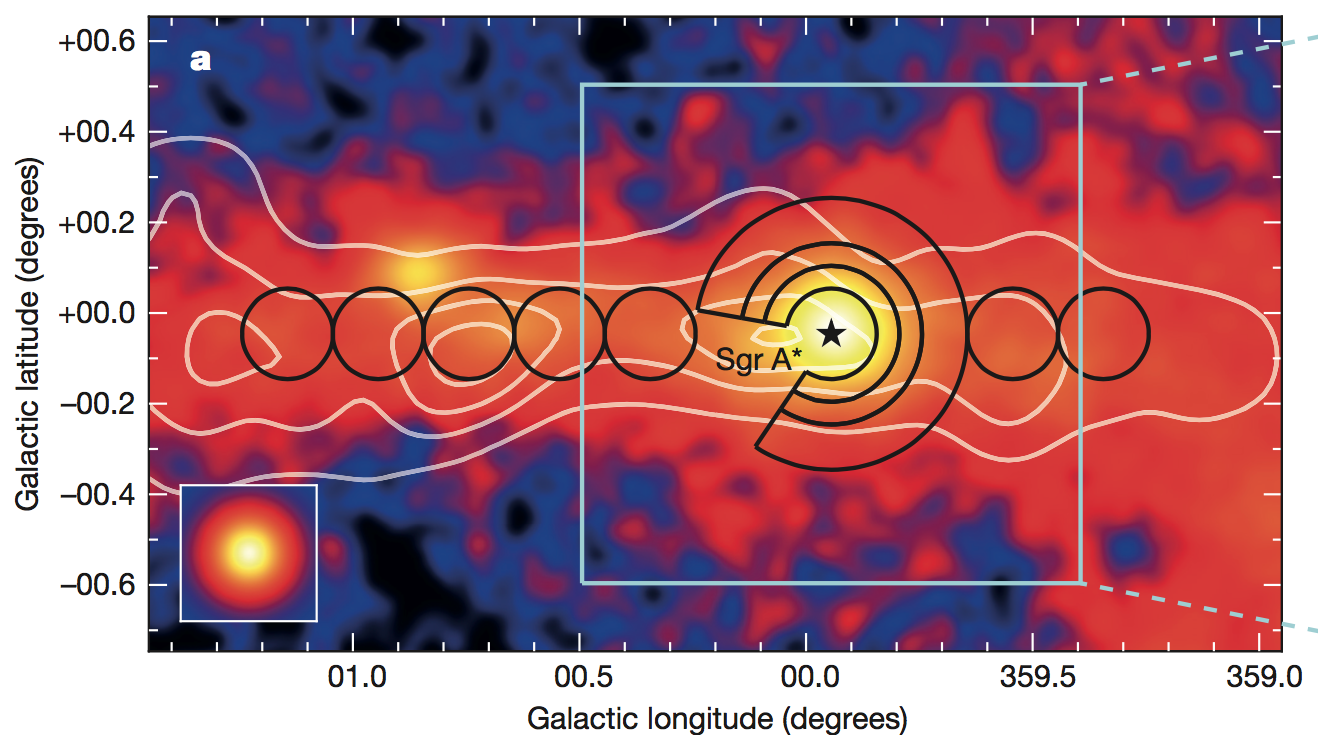}
\hspace{0.55cm}
\includegraphics[width=0.38\textwidth]{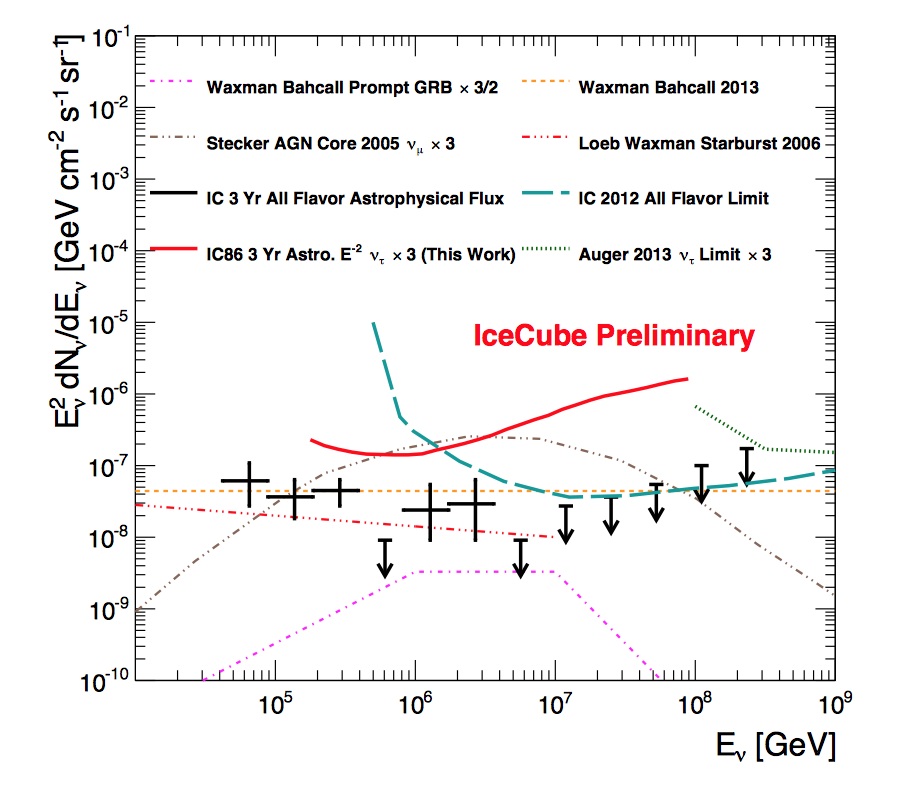}
\caption{Right:$\gamma$-ray image of the Galactic Centre region (see~\cite{HESSPevatron} for details). Right: Neutrino flux upper limits and models as a function of energy. The all flavor astrophysical neutrino flux observed by IceCube is shown as crosses~\cite{IceCubeNeutrino1}.  \label{fig:GammaRays}}
\end{figure}

At low energy, below 10$^{15}$~eV, a large number of galactic and extragalactic sources (see for example~\cite{Fermi-LAT2,Fermi-LAT3}) have been identified with satellite-based experiments and with  Imaging Air-Cherenkov Telescopes (IACT). The angular and energy resolutions achieved by the new instruments allow for the characterization of different classes of sources. An example is the Crab-nebula whose spectrum and pulsed emission have been extensively studied~\cite{HESScrab,MAGICpulse,VERITASpulse}. 
 A Galactic plane survey has been performed up to energies in the TeV range~\cite{HESSgalacticplane} and the evidence 
of a PeVatron host in the center of our Galaxy (Fig.~\ref{fig:GammaRays}, left) has been recently reported~\cite{HESSPevatron}.  
Moreover, in the past year the Fermi-LAT experiment discovered two giant lobes of $\gamma$-ray emission extending above and below the Galactic Center, named Fermi bubbles, which could indicate a cosmic-ray acceleration near the Galactic Center~\cite{FermiBubbles}. 
Because of the interaction on the background radiation, the observation of extragalactic sources and of the diffuse $\gamma$-ray flux start to constrain the extragalactic background light (EBL) and the nature of cosmic-rays at the highest energies (see for example~\cite{EBLHESS,Liu,Gavish,BerezinskyDiffuseGamma}). Surface array detectors, having a wide field-of-view and a 100\% duty cycle, complement the IACT, especially for the monitoring of transient sources  and energy ranges up to PeV~\cite{HAWC,Tibet-AS,Argo-YBJ} but with worst angular and energy resolutions. 
 A review of some up-to-date results and of the sensitivity of the next-generation experiments can be found in~\cite{reviewGammaDirect,reviewGammaIndirect,CTA}. 
 
 Complementary to $\gamma$-rays, a flux of high energy neutrinos have been finally observed by the IceCube experiment~\cite{IceCubeNeutrino1} with energies up to the PeV. This  measurement is of particular interest because the observation of astrophysical neutrino can provide insights on the identification of cosmic-ray sources at the highest energies and on the understanding of the acceleration mechanisms since they are produced in hadronic processes.  The distribution of arrival directions of events observed by IceCube does not yield significant evidence of clustering~\cite{IceCubeDirection}. Several search of neutrinos from point-sources have been performed also with combined analysis with the ANTARES underwater neutrino experiment~\cite{ANTARES} and other ultra-high-energy cosmic rays observatories~\cite{IceCubeAuger}.  The point-source searches with the ANTARES detector are particularly interesting as its location in the Northern hemisphere allows for the observation of the galactic center region. Next-generation experiments (e.g., IceCube-Gen2, KM3Net-ARCA) aim to increase the instrumented ice/water volume and optimize the detectors~\cite{IceCubeGen2,KM3NetARCA}. Moreover they include also low energy extension mostly aiming at the study of neutrino masses~\cite{PINGU, ORCA}. 
It is important to underline the role that cosmic particles can play in the next years beyond the astrophysics and the standard physics model. More specifically, cosmic rays have demonstrated their potentiality to study dark matter particles with masses around hundreds of GeV or $>$TeV $-$~where the current direct search experiments have limited sensitivity~$-$ and to test fundamental physics (for example the axion-like particles, the Lorentz Invariance Violation, the sterile flavor neutrino~\cite{IndirectDM,FundamentalPhysics,FundamentalPhysicsCTA,sterileIceCube}). 

\section{Direct measurements: the GeV - TeV energy range}

At low energy, the large flux of low-energy cosmic rays allows their detection with space-based experiments. 
In addition to the energy spectrum for different chemical elements, the abundance of isotopes or secondary particles (produced by spallation in the interstellar medium, as lithium and borum) with respect to the primary cosmic-ray flux  are measured to estimate the average amount of material that cosmic-rays typically traverse before reaching the observer~\cite{BESSiso,PAMELAiso,BC}.  
A review of the status of the spectral and mass composition measurement up to 10$^{12}$~eV can be found in~\cite{PDG}. 
The energy spectra up to hundred GeV derived for the different masses can be well described by single power-laws: the measured spectral indices, within uncertainties, supporting the hypothesis of similar acceleration and propagation processes for all the components. In this energy range, the most abundant elements are proton and helium which constitute almost 98\% of the all-particle spectrum. 
\begin{figure}[!t]
\centering
\includegraphics[width=0.45\textwidth]{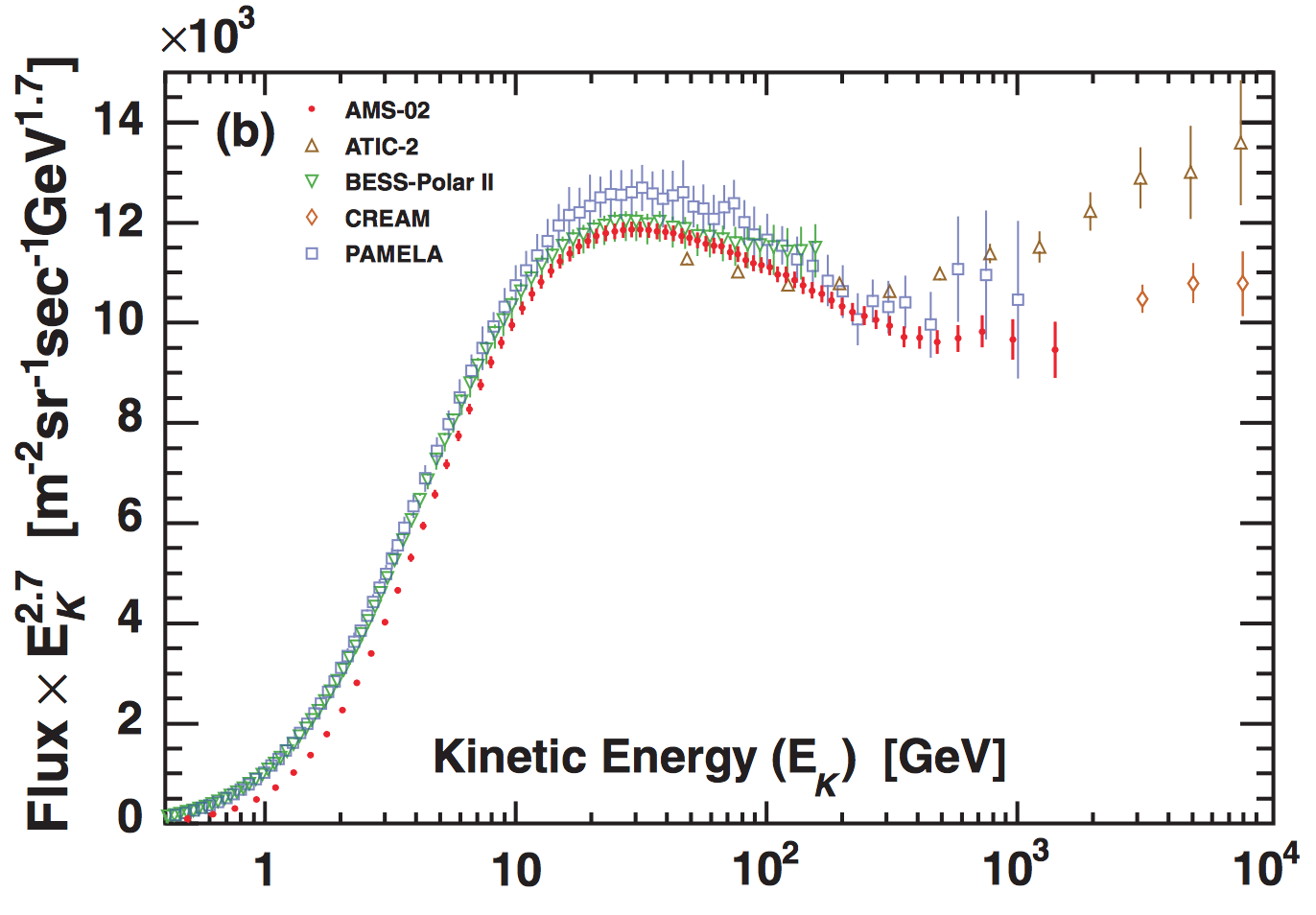}
\includegraphics[width=0.46\textwidth]{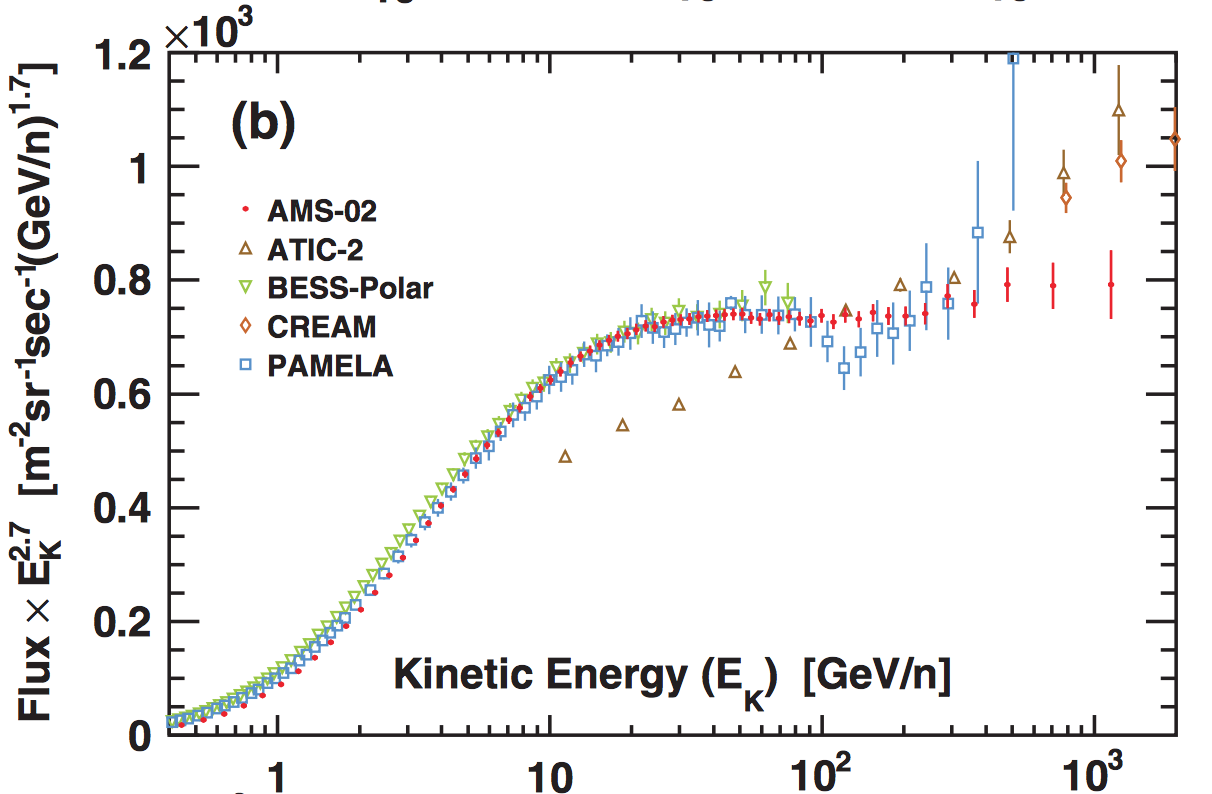} 
\caption{Proton (left) and helium (right) spectra measured by the AMS Collaboration and previous direct detection experiments (see~\cite{AMSp,AMSHe} and ref. therein).\label{fig:spectraGeV}}
\end{figure}
Their spectra (Fig.~\ref{fig:spectraGeV}), recently measured by the AMS-02, manifest a hardening at energies of about 100~GeV~\cite{AMSp, AMSHe, PAMELApHe} which can be explained as 
\begin{figure}[!t]
\centering
\includegraphics[width=0.47\textwidth]{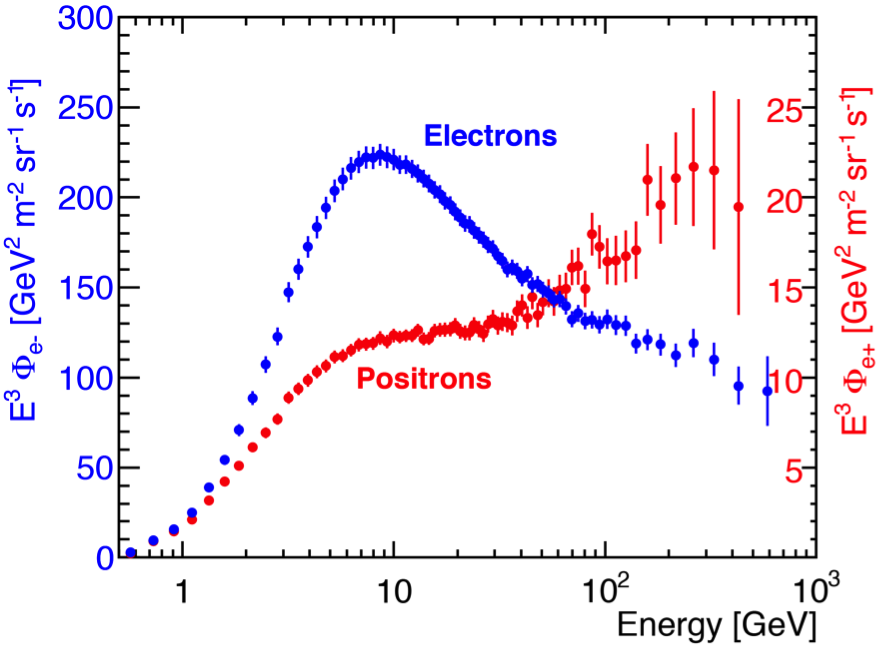}
\includegraphics[width=0.43\textwidth]{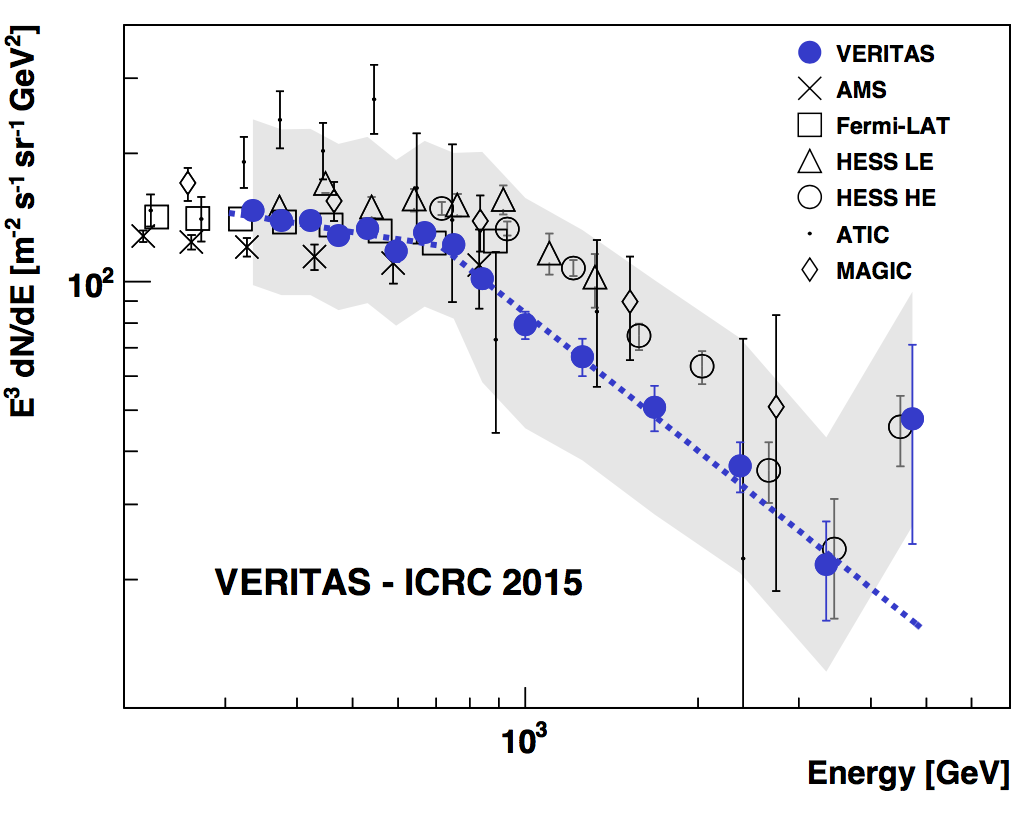}  
\caption{Left: positron and electron spectra measured by the AMS Collaboration up to 500~GeV. Right: electron spectra measured with the air-Cherenkov telescopes installed at ground~\cite{VERITASe,MAGICe,HESSe} and by satellite-based experiments~\cite{FERMIe, ATICe}\label{fig:e+e-}}
\end{figure}
propagation effects, re-acceleration processes or the existence of SNR contributing to different energy ranges (see~\cite{LowEnergyDirect} and ref. therein). 
About 1\% of the cosmic-ray flux consists of electrons and positrons which are expected to be produced during the propagation of cosmic rays (mostly from pion decay) and there are evidences of their acceleration in SNR shocks~\cite{electronsSNR}. In addition, they could originate from dark matter decay or annihilation and are therefore used for the indirect search of dark matter. Their flux has been measured by AMS-02~\cite{AMSe+e-,AMSe-,AMSe+}  in the energy ranges from 0.5 up to hundreds of GeV depending on the specific analysis (Fig.~\ref{fig:e+e-}, left). 
The positron excess initially reported by PAMELA~\cite{PAMELAe+,PAMELAe-} has been confirmed by  AMS-02: it  increases up to energies of about 25 GeV then becomes stable. It has been interpreted as a possible dark matter signature even if some models of production of primary positrons in astrophysical sources or the generation of an excess during cosmic-rays propagation can still describe the data. The improvement in the cosmic-ray propagation parameters would be important to reduce systematic uncertainties and eventually exclude some of these scenarios. 
At higher energies, the  measurement of the electron flux is only possible with IACT at ground and the current results indicate a flux cut-off at energies around 1~TeV.  
New experiments aiming at the measurement of the individual elemental spectra up to hundreds of TeV are foreseen in the future (e.g., NUCLEON, CALET, DAMPE, ISS-CREAM, GAMMA-400) and are currently at different stages of design, construction and operation. 

\begin{figure}[!t]
\centering
\includegraphics[width=0.53\textwidth]{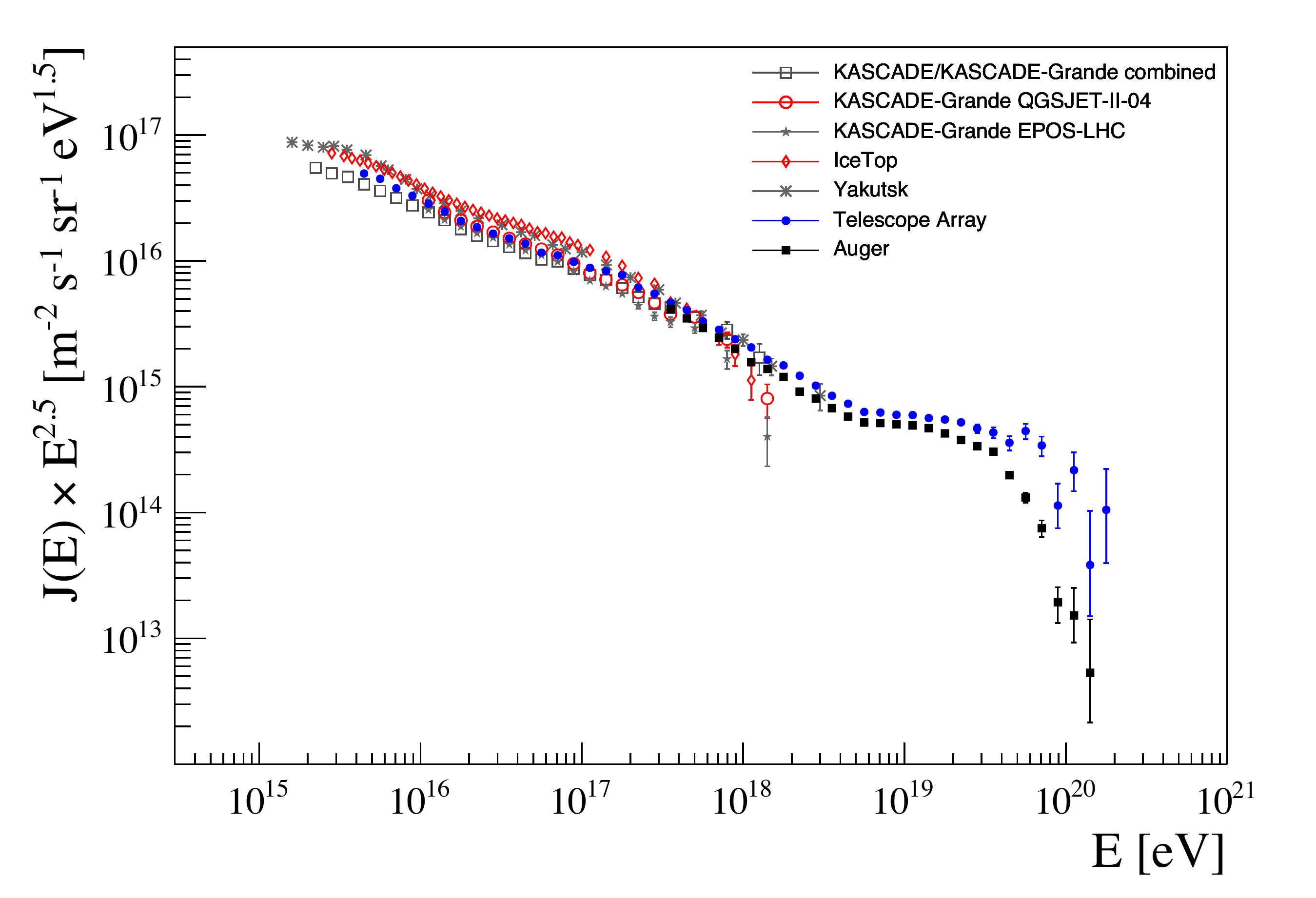}
\includegraphics[width=0.43\textwidth]{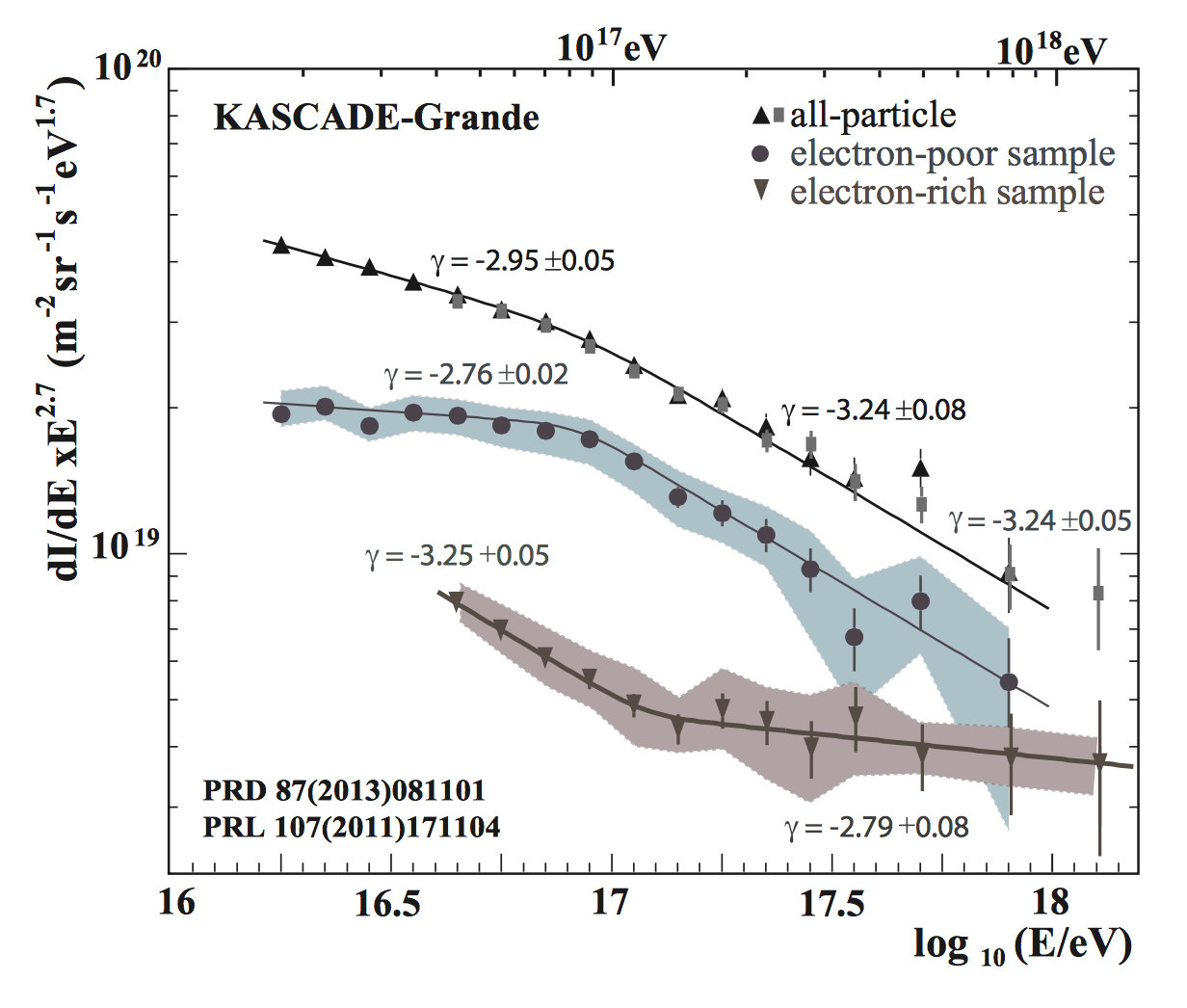}
\caption{Left: most up-to-date all-particle energy spectra presented in the energy range above 10$^{15}$~eV. Right: all-particle, electron-rich (light mass) and electron-poor (heavy mass) energy spectra from KASCADE-Grande. \label{fig:spectra}}
\end{figure}

\section{Indirect measurements: from the knee to the cut-off}

At energies above tens of TeV, cosmic rays are studied via the particle cascade they produce in atmosphere. Two techniques are mostly used: the observation of the density of particles at ground with an array of surface detectors (SD) and the observation of the longitudinal shower developed with fluorescence telescopes (FD). 
The first technique relies on the simulations of the air-showers for the calibration of the energy estimator.   
On the other hand, the FD allows for a calorimetric measurement of the energy and a direct observation of the shower-maximum depth ($X_{\rm{max}}$) which is a mass-composition sensitive parameter. However the FD suffers of a duty cycle limitation to about 10-15\%.
Several efforts to develop new detection techniques to observe the longitudinal profile without  the statistical limitation of the FD have been performed showing promising results up to 10$^{18}$~eV~\cite{Lofar1,aera}. \\

One open question at energies above the ``knee'' is the place of the transition between galactic and extragalactic component. 
A steepening of the spectrum, around a few times 10$^{16}$~eV, and a hardening above 10$^{17}$~eV have been reported by several experiments~\cite{Kascade, KascadeGrande,IceTop,Tunka133,Yakutsk,TALE}. The measured spectra are compatible within the systematic uncertainties and this is a remarkable result if considering the different detection techniques and the different hadronic-models and mass composition assumptions. The  KASCADE-Grande experiment has also shown that the steepening of the flux at 8$\cdot 10^{16}$~eV is emphasized when selecting only electron-poor events (heavy primaries) whereas the flattening above 10$^{17}$~eV is mostly related to electron-rich events (see Fig.~\ref{fig:spectra}, right). 
Whereas these spectral features can be fit with a rigidity model, the predominantly light composition observed around and above 10$^{17}$~eV challenges this interpretation~\cite{Lofar2,XmaxICRC}. 
Several models trying to describe the current measurement of the energy spectrum and the mass composition in the extragalactic-transition and ankle region have been proposed assuming a second (galactic) component, acceleration in compact sources and photo-disintegration in the acceleration ambient~\cite{Hillas25,Aloisio,Giacinti,Unger,Globus}.  \\
\begin{figure}[!t]
\centering
\includegraphics[width=0.48\textwidth]{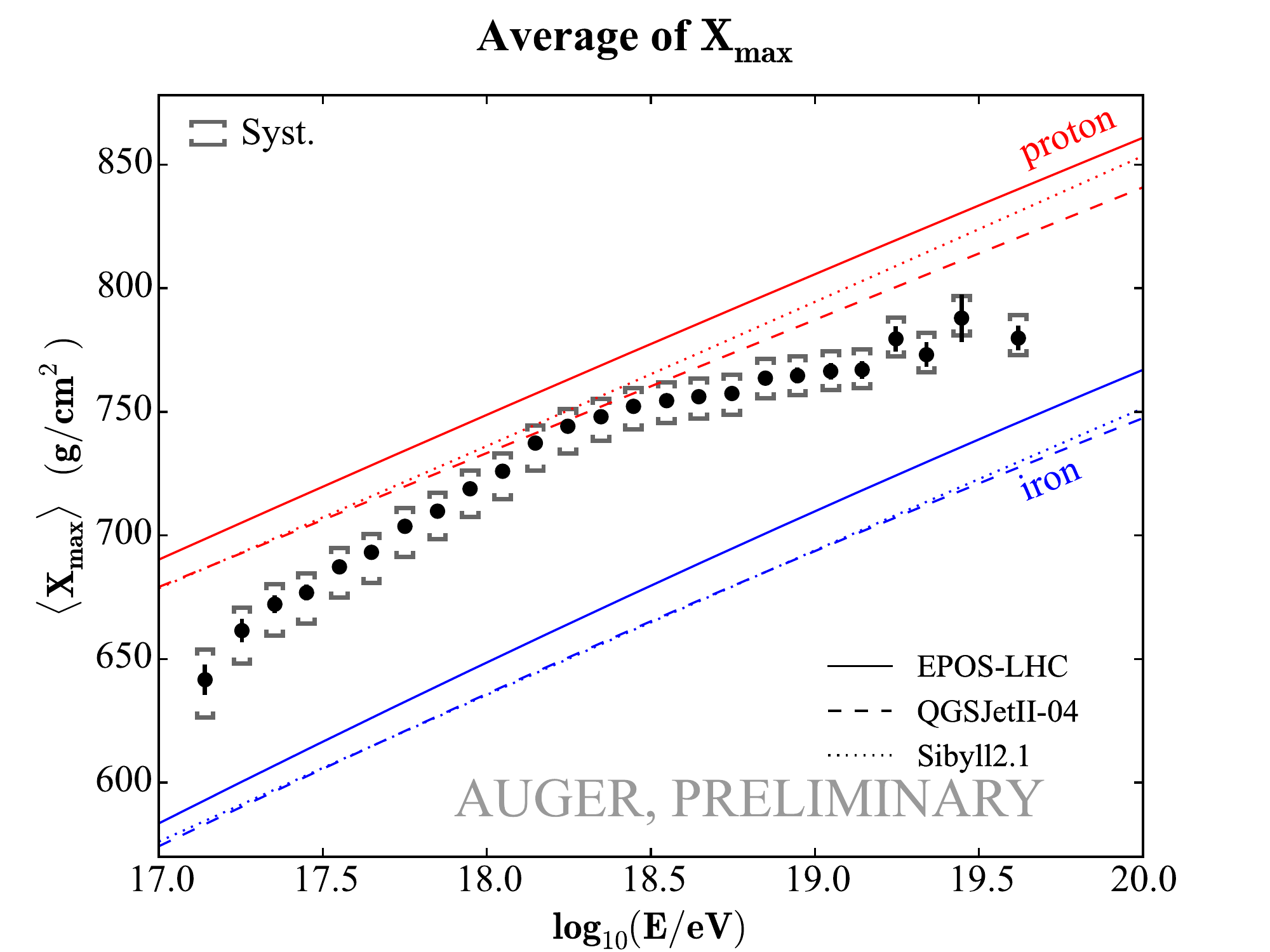}
\includegraphics[width=0.47\textwidth]{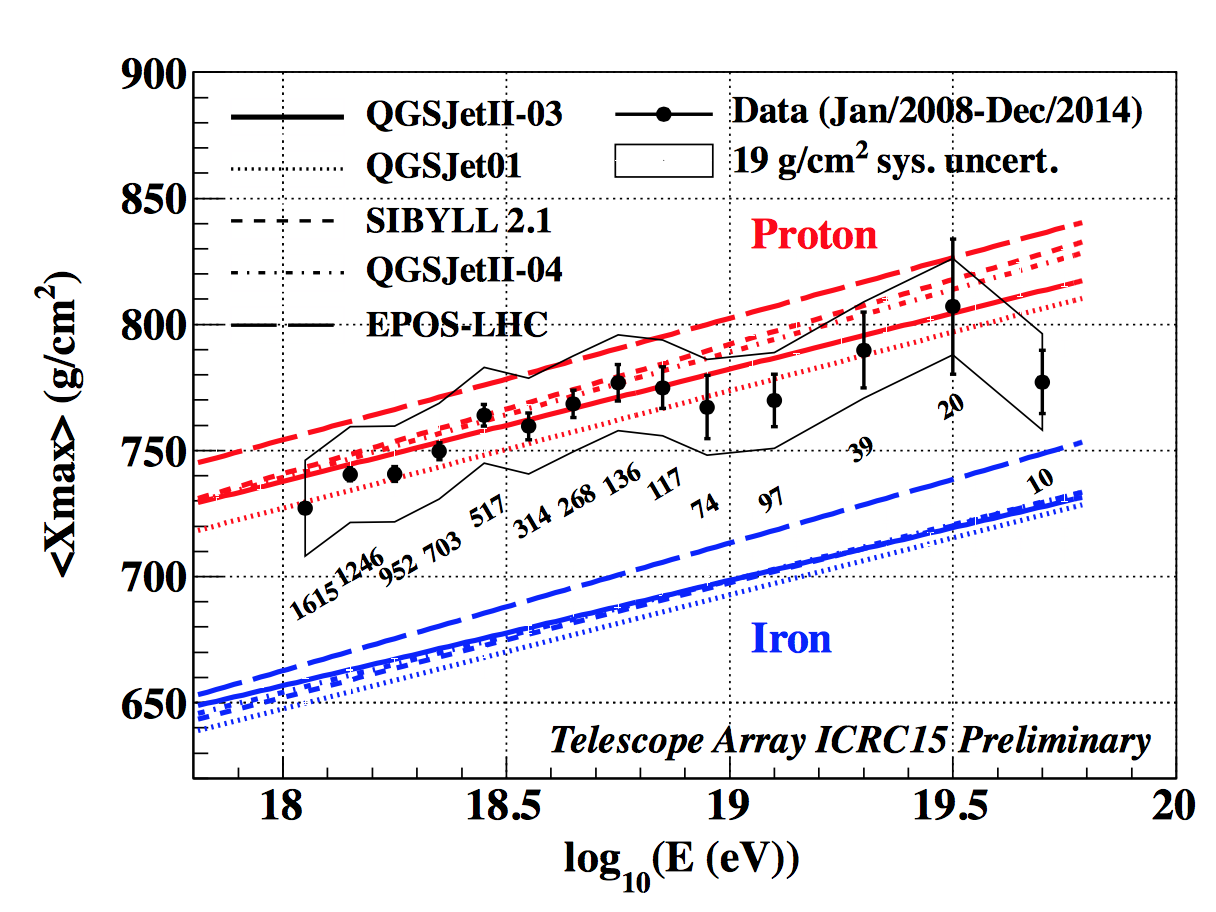}
\caption{\label{fig:Xmax}Evolution of the first moment of the $X_{\rm max}$ distribution with energy for the Auger (left) and Telescope Array (right) observatories. The expectations for proton and iron primaries are shown as lines for different hadronic interaction models. For the Auger Observatory selection criteria are applied to the data to remove the bias due to the limited field of view of the fluorescence telescopes. In the case no hard event selection is applied and the expectation lines are drawn after processing simulation with the full analysis chain to reproduce the same acceptance bias as in data.}
\end{figure}

At ultra-high-energies, above about 10$^{17}$~eV, two observatories  are currently taking data: the Telescope Array and the Pierre Auger Observatory in the northern and southern hemispheres, respectively. They employee an hybrid approach which combines the surface and fluorescence techniques to perform a data-driven calibration overcoming the limitations due to the FD duty cycle and the SD hadronic-model dependence. 
One of the most relevant results is the precise determination  of the ``ankle" position and the evidence for the flux suppression at energies above 10$^{19.5}$~eV.  
The energy spectra measured by the Pierre Auger and the Telescope Array observatories (Fig.~\ref{fig:spectra}, left) are compatible within the systematic uncertainties on the energy scale of the two experiments, quoted as 14\% for the Pierre Auger Observatory and 21\% for Telescope Array~\cite{AugerSpectrum2015,TASpectrum2015}. 
The possibility that the discrepancy in the cut-off region originates from a different sky observed in the northern and southern hemispheres has been tested: no evidence of a declination-dependence of the spectrum has been reported by the Auger Observatory whereas results from Telescope Array are not yet conclusive. The interpretation of the flux suppression is still in discussion.  
\begin{figure}[!t]
\centering
\includegraphics[width=0.53\textwidth]{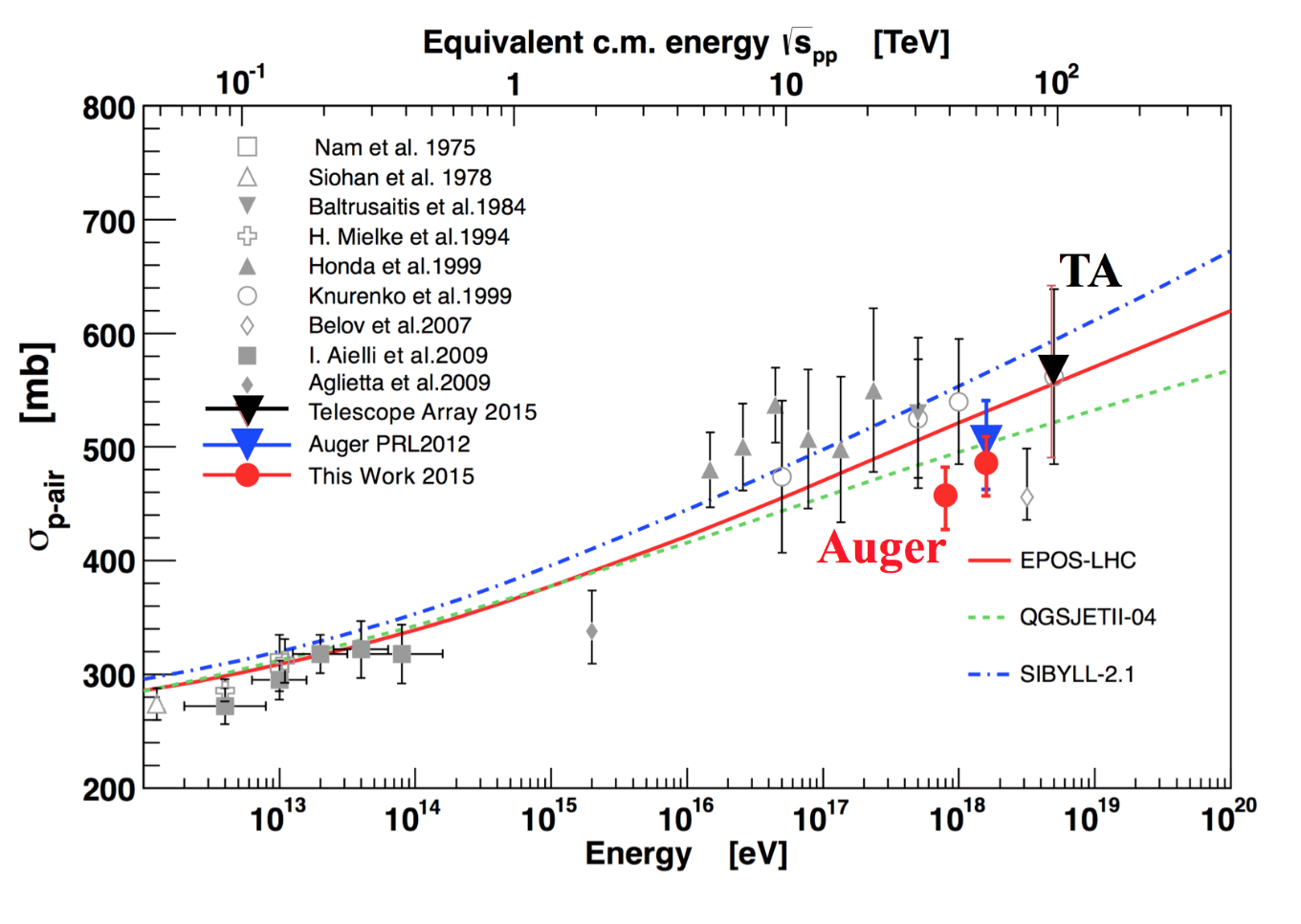}
\includegraphics[width=0.43\textwidth]{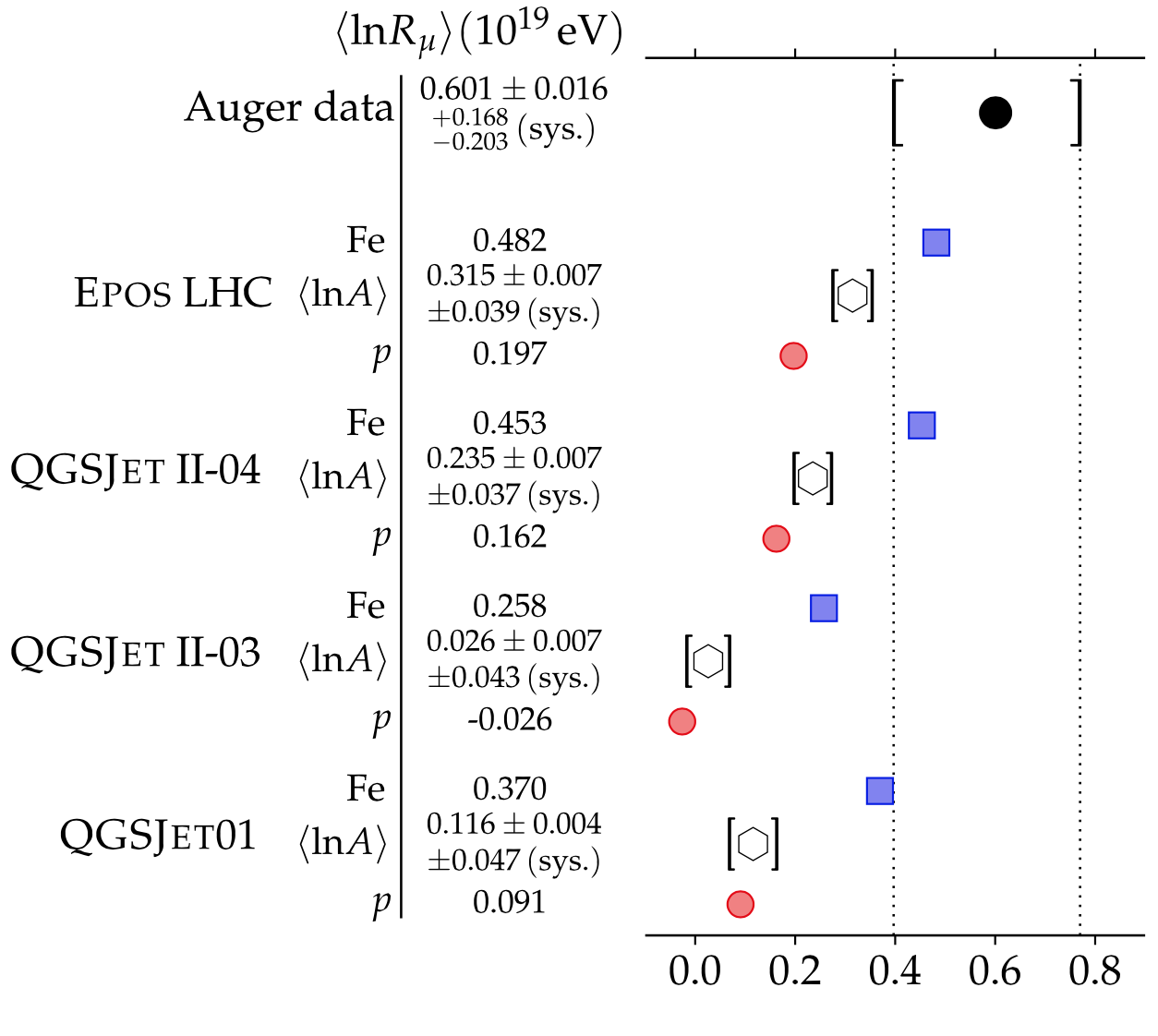}
\caption{Left: Measurement of the proton-air collision cross section from the Telescope Array and Auger observatories. Right: muon deficit observed comparing air-shower simulations and Auger data for different hadronic models and mass composition assumptions. 
\label{fig:hadronic}}
\end{figure}
The mass composition measurement above $10^{19.5}$~eV and the observation of cosmogenic neutrinos and photons can give hints to disentangle between the GZK cut-off and the source-exhaustion scenarios. In the multi-messenger approach, the recent results on cosmogenic neutrino flux obtained by IceCube~\cite{IceCubeNeutrino2} and by the diffuse $\gamma$-radiation measured by Fermi-LAT~\cite{Liu, Gavish,BerezinskyDiffuseGamma} start to constrain the parameters of the proton-dominated scenario. 

The mass composition is currently inferred from the first first two moments of the $X_{\rm max}$ distribution ($\langle$$X_{\rm max}$$\rangle$ and $\sigma$($X_{\rm max}$)) which are related, via Monte Carlo techniques, to the ln$A$ and $\sigma$(ln$A$), respectively, with $A$ the atomic mass of the primary cosmic ray. However, because of the limited FD duty cycle, these measurements only extend up to 10$^{19.5}$~eV and cannot be extrapolated to the flux cut-off region. 
Results are shown in Fig.~\ref{fig:Xmax} for Auger (left) and Telescope Array (right)~\cite{XmaxAuger,XmaxTA}. The  composition evolves from mixed to light primaries at low energies and again to intermediate masses at energies larger than 10$^{18.3}$~eV.  Results from TA are interpreted by the collaboration as a pure proton over the full energy range. It is worth noting that the interpretation of these results depends on the hadronic models chosen as reference. A study reconstructing the mass mixture observed by the Pierre Auger Observatory using the full analysis chain of the Telescope Array collaboration showed that, within TA uncertainties, the two scenarios, Auger reconstructed mixture or pure proton, cannot be discriminated. 
It has also been shown~\cite{XmaxInterpretation} that the two first moments of an $X_{\rm max}$ distribution can be described with different mass composition mixtures. To avoid this degeneracy the full $X_{\rm max}$ distribution is fitted with templates obtained with simulations assuming a mixture of $N$-components whose abundances are free parameters of the fit. The best description of the data is obtained with four components (proton, helium, nitrogen and iron nuclei) and it shows a large fraction of protons at energies around the ankle and a negligible iron content over the full energy range, independently of the hadronic models.

In connection with the mass composition interpretation, it is worthwhile to mention the possibility to test hadronic physics with air-showers at center-mass energies that are one or two orders of magnitude higher than the ones reached at LHC. A measurement of the proton-air cross-section has been performed by the Pierre Auger and Telescope Array collaborations from the fit of the tail of the $X_{\rm max}$ distribution for a  sample of proton-dominated events. To select a proton-dominated dataset the Auger collaboration uses an energy range around 10$^{18.5}$~eV whereas for Telescope Array the full energy range is used~\cite{xsecAuger,xsecTA}. Results are shown in Fig.~\ref{fig:hadronic}, left. 

For the Auger Observatory, the surface detector has also been used to infer the muon content in the air-showers. 
This estimator, obtained from highly-inclined events, suggests a muon deficit in the air-shower simulations which may varies from 30\% to 80\% depending on the hadronic models (Fig.~\ref{fig:hadronic}, right)~\cite{NmuInclined}. This result is also confirmed from an independent analysis using hybrid events with zenith angle smaller than 60$^\circ$~\cite{NmuVertical}.  \\

\begin{figure}[!t]
\centering
\includegraphics[width=0.42\textwidth]{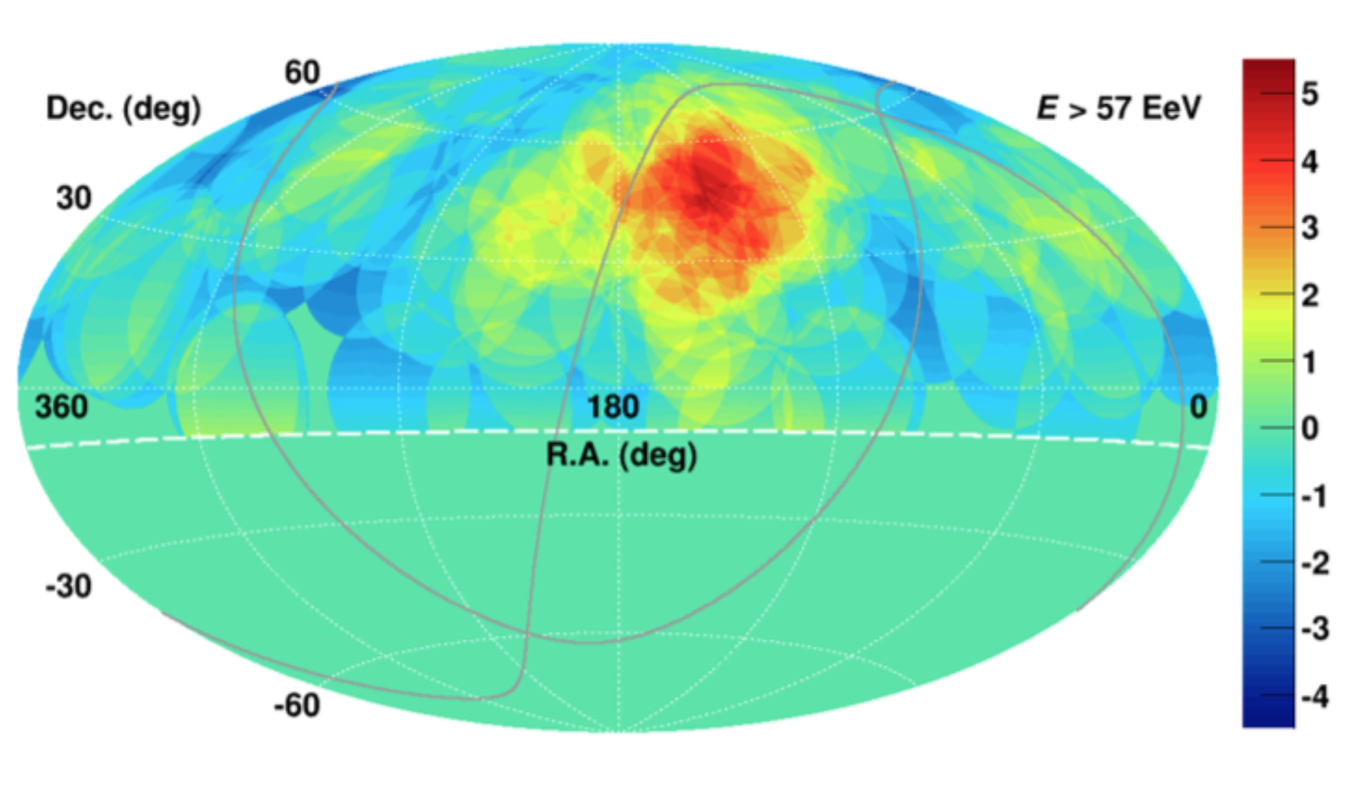}
\hspace{0.3cm}
\includegraphics[width=0.52\textwidth]{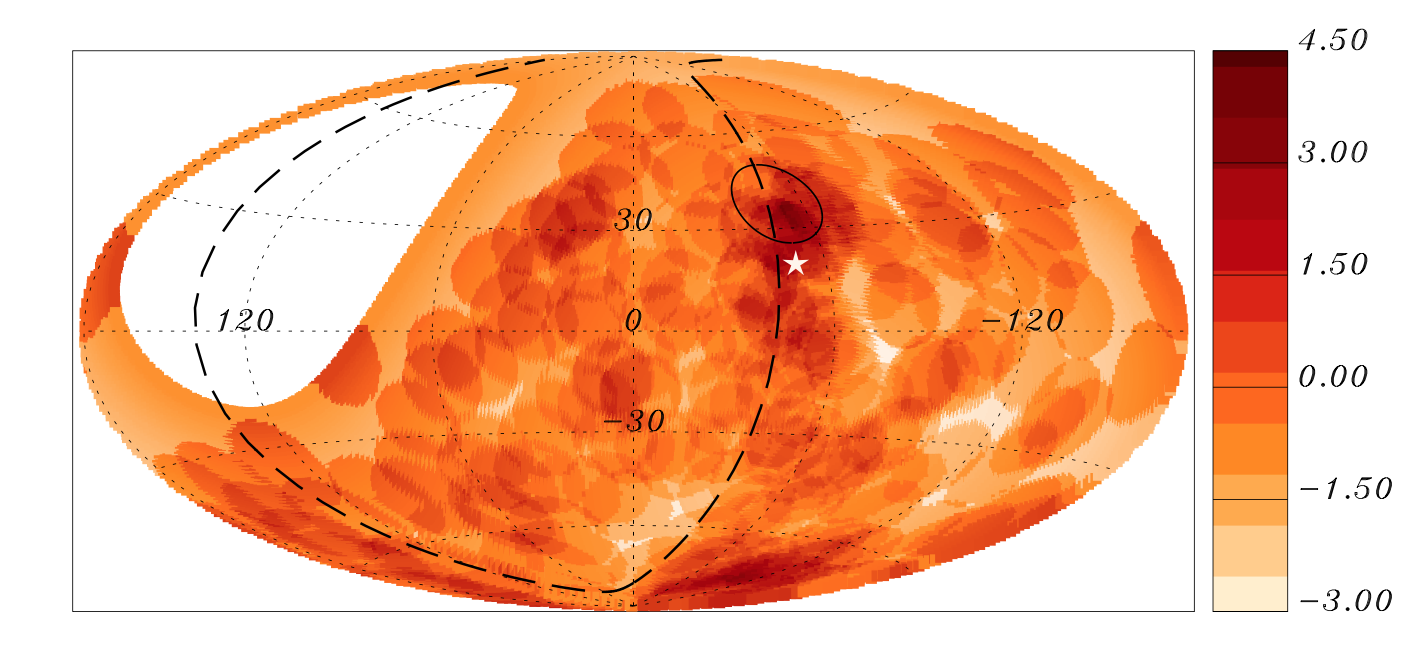}
\caption{Sky maps of the arrival directions of cosmic rays at the highest energies from Auger and Telescope Array. The colored zones indicate the field of view of each observatory. Left: The largest excess has been found by Telescope Array with a 3.4$\sigma$ (post-trial) for an energy threshold of 57~EeV and at an angular scale of 20$^\circ$. Right: map of a blind search for the events recorded by Auger. No significant deviation from isotropy is found and the most significant excess is obtained for events  above 54~EeV at 12$^\circ$ angular scale around the position of Centaurus A (white star).  
 \label{fig:anisotropy}}
\end{figure}

The possibility of doing astronomy with cosmic-rays above 10$^{19.5}$~eV has been considered feasible for a long time under the hypothesis of a proton composition and not-strong extragalactic magnetic fields ($\lesssim$ nG). The most up-to-date results show a surprisingly weak anisotropy signal at any angular scale.  
The most significant deviation from isotropy is found for events with energies above 57~EeV at an angular scale of 20$^\circ$ for Telescope Array~\cite{HotSpotICRC2015} with a post-trial significance of 3.4$\sigma$. For the Pierre Auger Observatory the largest  excess, not statistically significant, is found in the direction of Centaurus A ($E > 55$~EeV and angular scale of 15$^\circ$). At lower energies the search for anisotropy at large angular scales is also weak~\cite{LargeScale1,LargeScale2}. 
Unless assuming very strong magnetic fields, the dipole and quadrupole amplitudes obtained between 1 and 20 EeV constrain the galactic proton fraction to less than 10\%,  if sources are stationary and uniformly distributed in the galactic disk. Upper limits are otherwise compatible with a scenario with cosmic rays having a heavier composition of galactic origin and light elements of extragalactic origin~\cite{LargeScaleInterpretation}. \\

Many results have been obtained but a coherent picture is still missing. Answering the questions about the origin of the flux cut-off, the mass composition and the anisotropy of the arrival directions at the highest energies and constraining the hadronic interaction models are among the major outcomes expected from the upgrade of the Telescope Array and the Pierre Auger observatories. The Telescope Array upgrade (named ``TAx4") will extend the covered surface by a factor four, for a total of about 3000 km$^2$ as the Auger surface array, with the main goals of increasing the event statistics in the flux cut-off energy region and of testing the observed hot-spot with higher significance~\cite{TAx4}. 
The AugerPrime upgrade aims to perform a mass composition study using the surface detector (duty cycle about 10 times larger than the FD one) measuring separately the electromagnetic and muonic component at the ground. For this purpose, it will consist of a scintillator installed on top of each surface station, an improvement of the SD electronics, the extension of the operation time for the FD and the completion of the underground muon detector~\cite{AugerPrime}.

\end{document}